\documentstyle[12pt,epsf,a4]{article}
\newcommand{\koppl}[3]{ \left[ {#1}\otimes{#2}\right]^{[#3]} }
\newcommand{\ket}[1]{\left|{#1}\right\rangle}
\newcommand{\braket}[2]{\left\langle{#1}\right.\left|{#2}\right\rangle}
\newcommand{\bramket}[3]{\left\langle\,{#1}\,\left|\,{#2}\,
            \right|\,{#3}\,\right\rangle}
\newcommand{\bqn}{\begin{eqnarray}}
\newcommand{\eqn}{\end{eqnarray}}
\newcommand{\beq}{\begin{equation}}
\newcommand{\eeq}{\end{equation}}
\newcommand{\x}{\times}
\newcommand{\CL}{{\cal L}}
\newcommand{\ovl}[1]{\overline{#1}}
\newcommand{\bm}[1]{\bibitem{#1}}
\begin{document}

\title{\bf Quark structure and weak decays of heavy mesons}

\vspace{2cm}
\author{\large Stefan Resag and Michael Beyer\\[2ex]
{\em Institut f\"ur Theoretische Kernphysik der Universit\"at Bonn,}\\
{\em Nussallee 14-16, 53115 Bonn, FRG}\\
{\sf FAX 0228-732328, Email:beyer@itkp.uni-bonn.de}
\\[11cm]
{\sf BONN TK-93-18}\\[-11cm]}
\date{}
\maketitle

{\bf Abstract:} We investigate the quark structure of $D$ and $B$
mesons in the framework of a constituent quark model. To this end, we
assume a scalar confining and a one gluon exchange (OGE) potential.
The parameters of the model are adopted to reproduce the meson mass
spectrum. From a fit to ARGUS and CLEO data on $B\rightarrow
D^*\ell\nu$ semileptonic decay we find for the Cabbibo Kobayashi
Maskawa matrix element $V_{cb}=(0.036\pm 0.003)\;
(1.32ps/\tau_B)^{1/2}$.  We compare our form factors to the pole
dominance hypothesis and the heavy quark limit. For non-leptonic
decays we utilize factorization and for $B\rightarrow D^{(*)}X$ decays
we find $a_1 = (0.96\pm 0.05)\; (0.036/V_{cb})$, and $a_2=(0.31\pm
0.03)$.

\newpage

\section{Introduction}

Heavy mesons provide a rich and exciting tool to investigate the
implications of fundamental interactions and symmetries. Numerous
questions related to the dynamics of strong interaction, determination
of standard model parameters, and the search for new physics have been
addressed in this context.

Considerable amount of effort has been put into the description of
mesons in terms of the underlying quark structure. For heavy mesons,
viz. charmonium and bottomonium, non-relativistic models have been
particularly successful in describing the mass
spectra~[1-11].  However, it has also been shown
that relativistic effects may be quite significant in the transition
observables of charmonium (and to some extent also in bottomonium),
and generally, agreement with experimental data is
improved~\cite{bey92}.  For a review see e.g.~\cite{lic87, luc91}.

Here, we extend an earlier approach using a constituent quark model
supplemented by relativistic corrections~\cite{bey92} to the case of
unequal mass constituents. We address the questions of i) strong
interaction dynamics to determine the mass spectrum, ii) relativistic
effects in heavy-heavy and heavy-light quark systems, iii) form
factors, and iv) semileptonic and non-leptonic decays.

The model is able to describe the meson mass spectrum for low
radiative excitations to a satisfactory degree. However, for light
mesons, it fails to reproduce the empirical decay observables.  In
particular for the $\pi$ meson decay parameters such as the pion decay
constant $f_\pi$, leptonic or $\gamma\gamma$ decays cannot be
described. This holds even with inclusion of relativistic
corrections~\cite{res92}, because the $\pi$ meson is treated as a
rather strongly bound $q\ovl{q}$ system of constituent quarks of mass
$m_q\simeq m_N/3$. 

Some progress for the light mesons might be expected from recent
developments using the Bethe-Salpeter equation for the $q\ovl{q}$
systems~\cite{lag92, her93, tie93, mue93}.  However, technical and
conceptional difficulties have restricted the use of the
Bethe-Salpeter equation to approximate treatments, so far.

To circumvent some of the problems connected to light mesons that
appear in non-leptonic decays, we follow the approach of Bauer, Stech and
Wirbel (BSW) and introduce empirical values for the decay constants,
if possible, to make our results less ambiguous~\cite{bau87}.

In addition, for non-leptonic decays we utilize factorization. It has been
proven quite useful, and seems to work reasonably well for
$B\rightarrow D^{(*)}X$ transitions~\cite{bau87}, since final state
interaction ($fsi$) effects might be negligible in such cases.  Also,
it has been shown for $D$ decays (at least for those that involve
only one form factor) proper inclusion of $fsi$ restores the validity
of factorization to the level of experimental accuracy~\cite{kam93}.

With the assumptions mentioned, we are now left with the determination
of the two form factors for $0^- \rightarrow 0^-$ transitions and the
four form factors for $0^- \rightarrow 1^-$ transitions. They will be
calculated in the frame work of our model, which will briefly be
surveyed in the next section.  We compare our results to the pole
dominance ansatz to describe the $q^2$ behavior of the form factors.

Recently, much attention has been paid to heavy quark effective theory
(see e.g.~\cite{neu93} and references therein), which relates
form factors of $B\rightarrow D $ to those of $B\rightarrow D^*$
transitions introducing heavy quark symmetries. All form factors are
then related to one universal function (i.e. Isgur Wise
function~\cite{isg90}). We will also compare our results to the
idealized limit of heavy quarks.

Beside the description of the model, we also give the meson mass
spectra in Sect.~2. In Sect.~3 we calculate form factors for
semileptonic decays, which we also use in Sect.~4 where we give our
results for non-leptonic decays. We summarize our main results and give
our conclusion in Sect.~5.

\section{The quark model}

In the constituent quark model presented here, mesons are treated as
quark-antiquark states, i.e. we do not consider any gluonic admixtures
e.g. $g{\ovl q}q$. Also no coupled channel effects are included
although these might be relevant in some cases~\cite{eic78}.
 
Confinement is modeled by a Lorentz scalar potential. As such it gives
rise to a Darwin and a Thomas precession term in a $(p/m)$
expansion. Furthermore, we assume that there is a residual short-range
quark interaction from one-gluon-exchange which leads to spin and
angular momentum dependent terms.

\begin{figure}[t]
\vspace{11cm}
\includegraphics{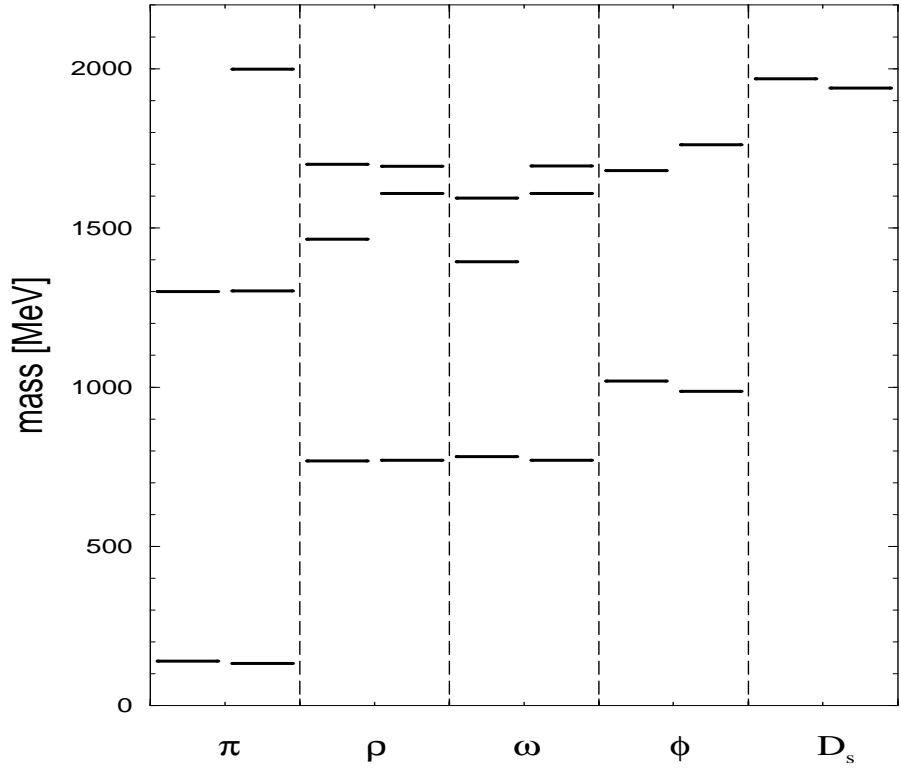}
\vspace{11cm}
\includegraphics{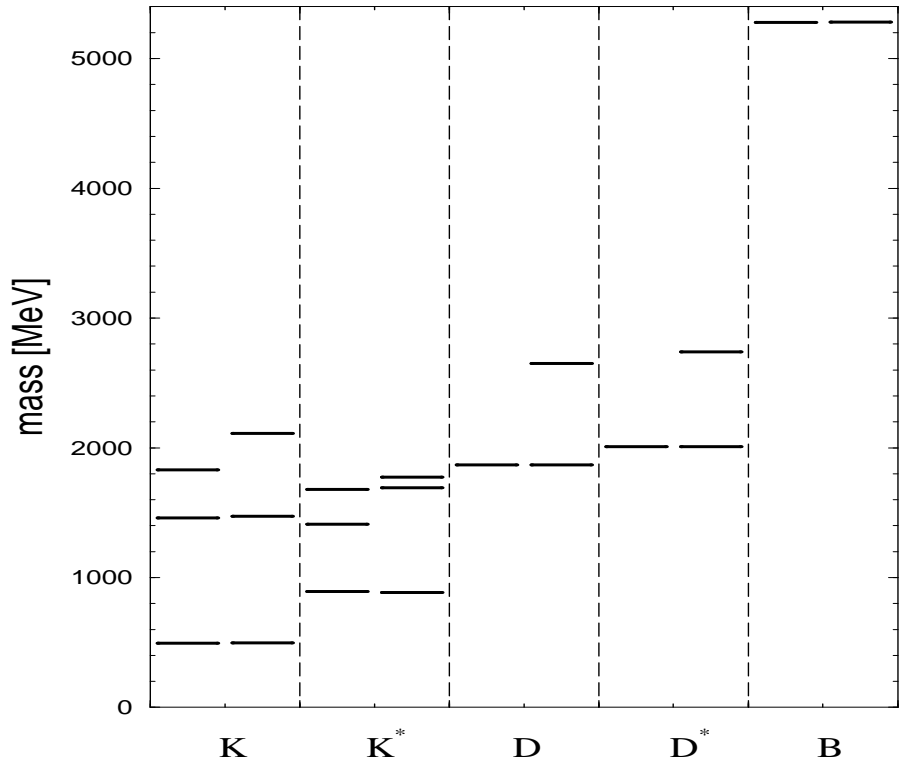}
\vspace{-1cm}
\caption{ \label{spectra}
 Meson mass spectra of pseudoscalar and vector mesons: 
  The l.h.s. of each column is the
  experimental data, while the r.h.s. shows the calculated values}
\end{figure}

However, in agreement with earlier results in heavy quarkonia, we
found that not all terms are equally important.  Differences between a
full version and a reduced version are rather small on the average
(see also discussion in~\cite{bey92}). For the low lying states of
$c{\ovl s}$, $b {\ovl s}$, $c{\ovl b}$, this conclusion is in
qualitative agreement with the findings of Lichtenberg, Roncaglia and
Wills who investigated the influence of different potential models in
these systems~\cite{lic89}.  In addition due to the mesons considered
here, viz. $0^-$ and $1^-$ mesons only, not all ingrediences of the
Hamiltonian contribute due to selection rules.  In particular
spin-orbit interactions do not contribute in the model space
chosen here. We define
\beq 
H=M+T+V_C + V_R + W_R^T+W_R^{SS}
\label{hamilton}
\eeq 
where $M$ is the
sum of the constituent quark masses $m_q$, $m_{\ovl q}$, and $T$ the
kinetic energy of relative motion in the center of mass system, 
\bqn
V_C&=&a+br\\
V_R&=&-\frac{4}{3}\frac{\alpha_s}{r} 
\eqn 
and the spin dependent forces
\bqn
\\ 
W_R^T&= &\frac{1}{3m_qm_{\bar{q}}}S_{q{\ovl q}}
                   \left( \frac{1}{r} {V_R^C}^\prime (r)-
                   {V_R^{C}}^{\prime\prime} (r) \right)\\
W_R^{SS}&= &\frac{2}{3m_qm_{\bar{q}}}
                   \left( {\bf s}_q \cdot {\bf s}_{\bar{q}} \right)
                   \left( \Delta V_R^C \right) (r)
\eqn 
In the above expressions, {\bf r} denotes the relative distance
between quark and antiquark, ${\bf s}_q$ and ${\bf s}_{\ovl q}$ are
the respective spins, ${\bf S}={\bf s}_q+{\bf s}_{\ovl q}$ the total
spin operator, and $S_{q{\ovl q}}=3{\bf s}_q\cdot\hat{{\bf r}} {\bf
  s}_{\ovl q}\cdot\hat{{\bf r}}-{\bf s}_q\cdot{\bf s}_{\ovl q}$ the
tensor operator ($\hat{\bf r}={\bf r}/r$). The quark masses and the
parameters of the potentials, the off-set $a$, the string tension $b$,
and the coupling strength $\alpha_s$ are treated as free parameters.
They are adjusted to reproduce the experimental meson spectra shown in
Fig.~\ref{spectra}. Parameter values
are shown in Table~\ref{parameter}.

We note here, that since we do not use a perturbative treatment of the
residual interaction, we need to regularize the potentials in order to
prevent presumably spurious~\cite{mcc83} divergences, see also
\cite{bey92}. We choose
\beq
1/r \rightarrow \sum_{i=1}^5 \beta_i \exp[-\gamma_i^2r^2]
\eeq
for the residual interaction, and
\beq
b\rightarrow b(r)=b\cdot\left(1-\exp[-(r/2r_0)^2]\right)
\eeq
for the confining interaction. The parameters $\beta_i$, $\gamma_i$
are fixed to fit $1/r$ with maximum likelihood in the region between
$r_0$ and $4r_0$, where $r_0$ is given in Table~\ref{parameter}.

The regularized Hamiltonian is then diagonalized in a reasonably large
basis. We have used a Laguerre (in momentum space) as well as an
oscillator basis, and found only marginal differences in the spectra.
In fact, results presented are calculated using the Laguerre basis,
with the advantages that the number of necessary basis states is
smaller and that the treatment of higher relative momenta is more
realistic due to the asymptotic behaviour of the Laguerre polynomials.
This might be of particular importance for small values of four
momentum transfers $q^2$, viz.  large three momentum ${\bf q}^2$ in
the rest frame of the decaying particle.

The energy eigenvalues are then obtained by minimizing the expectation
value with respect to the basis parameter due to Ritz' variational
principle. The minima occuring are rather flat, thus independent of
the specific choice of the variational parameter. Therefore it is possible
to choose the same parameter for all mesons.

The spectra for $0^-$ and $1^-$ mesons are shown in
Fig.~\ref{spectra}.  We have excluded all other mesons from the
figure, since they are not relevant for the present purpose.  However,
they have been included in the fitting procedure that lead to the
parameter values given in Table~\ref{parameter}.  To this end all
pieces of the Hamiltonian, such as e.g. the spin-orbit interaction
(but no additional parameters), have been retained~\cite{boh89}.  The
overall agreement of the model with the experimental values is quite
satisfactory.

The masses found for the constituent quarks are in agreement with the
values found by~\cite{lic89} for five different potential models. Also
our value for the off-set $a$ is close to the respective values of
Cornell~\cite{eic78}, Song-Lin~\cite{son87}, Turin~\cite{lic89}, and
Indiana~\cite{fog79} potential, also studied by~\cite{lic90}. The
value for $b$ is slightly smaller and $\alpha_s$ slightly larger than
the respective values of the Cornell potential, which has the same
$r$-dependence of $V_C$ and $V_R$.  Part of the differences may be
attributed to differences in the spin dependent part of the
interaction, but we do not go into further details here.

The different values of $r_0$ introduced to regularize the potential
due to different constituent quark masses $m_q$ quoted in
Table~\ref{parameter}, are not surprising and can be understood
qualitatively.  It may be interpreted as an effective extention of the
constituent quark (which in fact may be a complicated object) or
through the ``smearing'' of the potential due to relativistic
effects~\cite{god85}.

\section{Form factors and semileptonic decays}

Semileptonic decays are treated in current-current approximation.
Since we are mainly concerned about $b\rightarrow c$ transitions we
give all formulas for that case only.  Generalisation to different
flavor dependence is straight forward. The Lagrangian is then given by 
\beq
\label{lagrangian}
\CL_{cb}=\frac{G_F}{\sqrt{2}}\;V_{cb}\;\;h_{cb}^\mu\;j_\mu
\eeq
with the Cabbibo-Kobayashi-Maskawa matrix element $V_{cb}$.
The leptonic  $j_\mu$ and hadronic currents $h_{cb}^\mu$ are
defined by 
\bqn
j_\mu&= &{\ovl \ell} \gamma_\mu (1-\gamma_5) \nu_\ell\\
h_{cb}^\mu&=&{\ovl c} \gamma^\mu (1-\gamma_5) b
\label{curhad}
\eqn
The relevant transition amplitudes for $B\rightarrow D$ and
$B\rightarrow D^*$ of the hadronic current can be decomposed due to
the Lorentz covariance of the current, thus introducing form factors. For
$0^- \rightarrow 0^-$ transitions we use
\bqn
\nonumber
\lefteqn{\bramket{D, P_D}{h^\mu_{cb}}{B, P_B}  =} \\
&&\left( P_B + P_D -\frac{m_B^2 -m_D^2}{q^2} q\right)^\mu  F_1(q^2)
+  \frac{m_B^2 - m_D^2}{q^2} q^\mu  F_0(q^2)
\label{curBD}
\eqn
with
$q^\mu = (P_B - P_D)^\mu$, and $F_0(0)  =  F_1(0)$.
In the case  of $0^- \rightarrow 1^-$, the appropriate current may
be parametrized as follows
\bqn
\nonumber
\lefteqn{\bramket{D^*,P_{D^*}\varepsilon}{h^\mu_{cb}}{B, P_B}=}\\
\nonumber
&&\frac{2}{m_B + m_{D^*}} \epsilon^\mu_{\nu\rho\sigma}
\varepsilon^{*\nu} P_B^\rho P_{D^*}^\sigma
 \, V(q^2)
- i\varepsilon^* \cdot q  \frac{ 2 m_{D^*}}{q^2}\, q^\mu\, A_0(q^2)\\
\nonumber
&-& i (m_B + m_{D*}) \left( \varepsilon^*
-  \frac{\varepsilon^* \cdot q }{q^2}  q\right)^\mu A_1(q^2) \\
&+ &\frac{ i\varepsilon^* \cdot q }{m_B + m_{D^*}}
\left(P_B + P_{D^*} 
      - \frac{m_B^2 -m_{D^*}^2}{q^2}q\right)^\mu A_2(q^2)
\label{curBDS}
\eqn
with $\varepsilon^\nu$ polarization vector of $D^*$,
$\epsilon_{\mu\nu\rho\sigma}$ 
antisymmetric tensor, and the restriction
\beq
2m_{D^*}A_0(0) = (m_B + m_{D^*}) \, A_1(0)
\, - \, (m_B -  m_{D^*}) \, A_2(0) 
\eeq
Note that $0<q^2<q_{max}^2=(m_B-m_{D^{(*)}})^2$ due to kinematical
reasons. With the parametrization given above, it is straight forward to
evaluate observables. They have been given by K\"orner and Schuler in a
series of papers and formulas which will not be repeated here~\cite{koe88}.

Our concern is to determine the form factors. For special values, e.g.
at $q^2=0$ or $q^2=q_{max}^2$, the form factors may be fixed due to
(approximate) flavour symmetries and a conserved vector current.

It has been argued that at $q^2\simeq q^2_{max}$, i.e. small three
momentum transfer, values of the form factors may be fixed with a
nonrelativistic model.  This has been utilized by Godfrey and Isgur,
who use $m_q+m_{\ovl q}$ instead of experimental meson masses $\mu$
and connect their arguments to the Lorentz representation of free
particles~\cite{god85}.

\begin{figure}[t]
\vspace{20cm}
\includegraphics{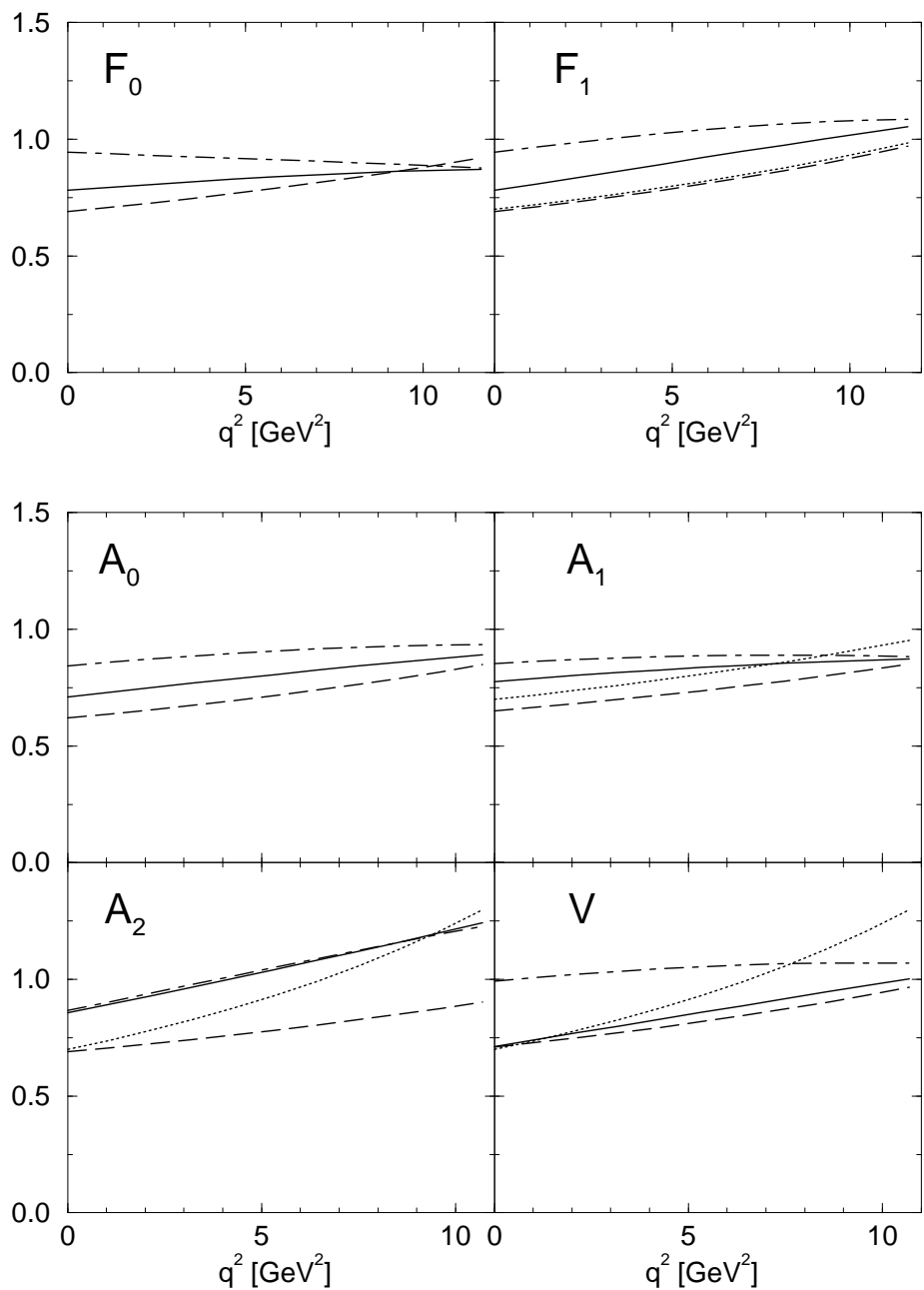}
\vspace{-1cm}
  \caption{\label{BDformfactor}
  A comparison of form factors for  B $\rightarrow$ D and
  B $\rightarrow$ $D^*$ transitions; our full result (solid line), 
  our non relativistic
  result (dashed-dotted), BSW (dashed), KS (dotted, where given)}
\end{figure}

A different possibility is to assume pole dominance. Such poles may
occur, if strong interaction processes dominate the decay dynamics,
and poles are formed prior to decay. This has been widely used by
Bauer, Stech and Wirbel (BSW)~\cite{bau87}, and also K\"orner and
Schuler (KS)~\cite{koe88} but with different pole types and different
values at $q^2=0$. BSW use ad hoc wave functions of the relativistic
harmonic oscillator (in the infinite momentum frame) to fix the value
at $q^2=0$.  KS assume all ($B\rightarrow D^{(*)}$) form factors
normalized to $0.7$ at $q^2=0$. Both models are compared to our
results in the following, see Fig.~\ref{BDformfactor}.

In order to evaluate the current matrix elements (i.e. l.h.s. of
(\ref{curBD}) and (\ref{curBDS})) from a quark model to determine the
form factors, we introduce a Fock-space representation for mesons.
This has been done e.g. by van Royen and Weisskopf~\cite{roy67} and by
Godfrey and Isgur~\cite{god85}. We follow their suggestion with minor
changes. This procedure allows us to introduce relativistic corrections
in the current operators as will be explained now.

The meson is represented as a superposition of free quark states,
and the amplitudes are given in terms of the meson wave function in
momentum space, viz.
\bqn
\ket{\omega,{\bf P}} &= &\sqrt{2\omega}\int\frac{d^3p}{(2\pi)^3}
\frac{1}{\sqrt{2p^0_q 2p^0_{\ovl q}}}\label{wavefct}\nonumber\\
&&\x\sum_{LS} R_{NLS}(p)
\koppl{Y_L(\hat {\bf p})}{\chi_S}{J}\chi_F\chi_C
\ket{\frac{m_{ q}}{M} {\bf P}+{\bf p}}_q
\ket{\frac{m_{\ovl q}}{M} {\bf P}-{\bf p}}_{\ovl q}
\eqn

where ${\bf p}_q$, ${\bf p}_{\ovl q}$ are the quark momenta,
$p_q^0=\sqrt{{\bf p}_q^2+m_q^2}$ (and analogously for ${\ovl q}$). The
total momentum is given by ${\bf P}={\bf p}_q+ {\bf p}_{\ovl q}$, and
the relative momentum by $M{\bf p}=m_{\ovl q}{\bf p}_q -m_{ q}{\bf
  p}_{\ovl q}$.  The quarks are normalized with $\braket{{\bf
    p}_q'}{{\bf p}_{ q}}=(2\pi)^3 2p^0_q \delta({\bf p}_q'-{\bf
  p}_q)$. The spin, flavor, and color wave functions are denoted by
$\chi_S$, $\chi_F$, $\chi_C$, respectively.  The momentum space wave
function is given by $R_{NLS}(p)Y_L(\hat {\bf p})$. Using the
experimental meson mass $\mu$, the free energy of the meson is
$\omega=\sqrt{{\bf P}^2+\mu^2}$. Thus the normalization chosen in
(\ref{wavefct}) differs from~\cite{god85}, and is given by
$\braket{{\bf P}'}{{\bf P}}=(2\pi)^3 2\omega \delta({\bf P}'-{\bf
  P})$.

The current operator (\ref{curhad}), evaluated between the meson wave
function given in (\ref{wavefct}) lead to the following single quark
matrix elements, 
\bqn
\label{quarkcurrent}
\bramket{ p'_{q}}{h^{\mu}_{cb}}{ p_{q}} 
& = &(2\pi)^{4}\,
\delta^{(4)}(p_{q}\,-\,p'_{q}\, -\,q)\,
\bar u_{q'}(p'_{q})\,
\gamma^{\mu}\,(1-\gamma_{5})\,u_{q}( p_{q})
\eqn
This way it is possible to take into account (relativistic) effects
induced through the full Dirac quark spinor. In the following the
nonrelativistic result quoted is achieved by taking the lowest order
$p/m$ in the above matrix element, and neglecting the momentum
dependence in the normalization of (\ref{wavefct}).

We then evaluate the hadronic matrix elements using (\ref{wavefct}) in
the laboratory system.  Although this approach is not a covariant
formalism, it is a natural way to include relativistic effects into
the calculation of decays in a nonrelativistic quark model. For heavy
mesons this formalism should be appropriate since $p/m$ is small in
these mesons.

In Fig.~\ref{BDformfactor} we give the form factors relevant for the
transition of $B\rightarrow D$ and $B\rightarrow D^*$. They are
calculated using the wave functions consistent with the Hamiltonian
(\ref{hamilton}), and reproducing the mass spectra as shown in Fig.
\ref{spectra}. Our calculation is given by the solid line. The (mono)
pole dominance ansatz of BSW is shown as a dashed line.  The
parameters of BSW are given in Table~\ref{polmasses}. Schuler and
K\"orner use a single pole mass of $m_{pole}=6.34$GeV.  The form
factors $F_1$ and $A_1$ are assumed monopole, but opposite to BSW,
$A_2$ and $V$ are assumed dipole behaviour and shown as dotted line
\cite{koe88}. Our results differ strongly to the (di-) pole ansatz.

In general our result is larger than the BSW model, so that a smaller
$V_{cb}$ can be expected than has been found by~\cite{bau87}. Taking
only nonrelativistic terms as explained above we find even larger form
factors than the full calculation. They differ less at $q^2=q_{max}^2$
(as expected) but rather strongly at $q^2=0$. The form factor $A_2$ is
only slightly changed.  The nonrelativistic results are shown as
dashed-dotted curves in Fig.~\ref{BDformfactor}.

Since the model is not covariant, form factors usually
depend on the reference frame. As a test, we evaluated $0^-
\rightarrow 0^-$ transitions in the $D$ rest frame also. Differences
in the from factors are generally smaller than 10\%. At $q^2=0$ we
found differences of a few per cent only compared to the calculation in
the $B$ rest frame .

\begin{figure}[t]
\vspace{18cm}
\includegraphics{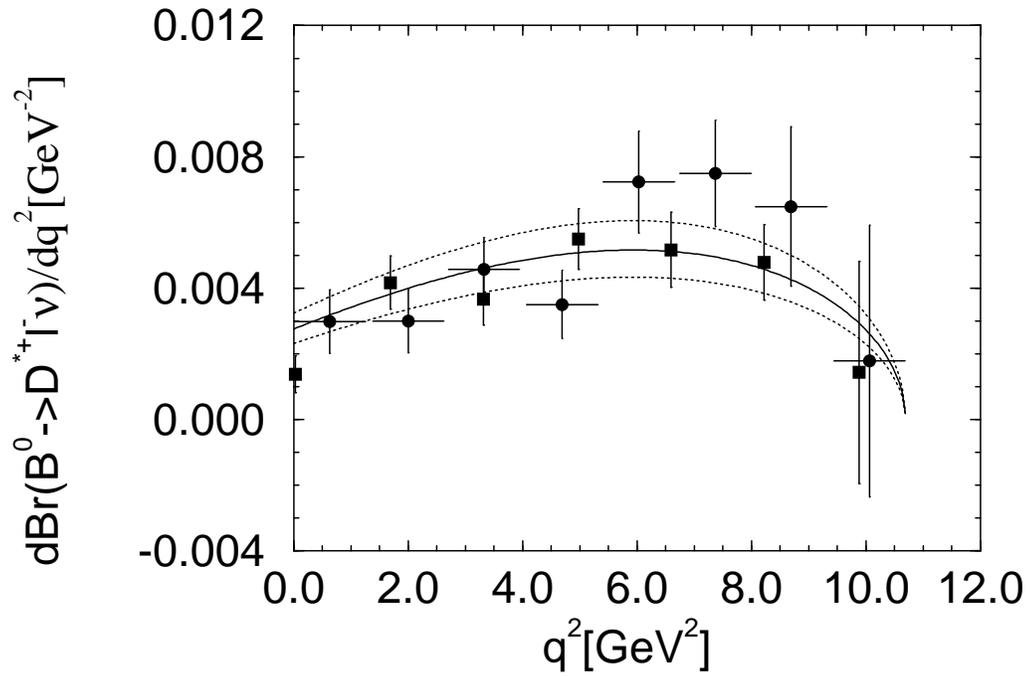}
\vspace{-8cm}
\caption{\label{decayBD}
  $q^2$ distribution of $B^0 \rightarrow D^{*+}\ell^-{\ovl \nu}$.
  Experiments given by ARGUS (circles) and CLEO (squares). 
  The solid line calculated with
  $V_{cb}=0.036$ and life time $\tau_B=1.32ps$; the upper and lower dotted
  lines with $V_{cb}=0.036\pm 0.003$ respectively.}
\end{figure}

The exclusive decay spectrum is shown in Fig.~\ref{decayBD}. We
compare our results to recent ARGUS~\cite{alb93} and CLEO data
\cite{cas93,san93}. We find a best fit with $V_{cb}=(0.036\pm 0.003)
(1.32ps/\tau_B)^{1/2}$.  The error resulting from $\chi^2$ fitting is
indicated by the upper and lower dotted line.

The resulting total branching ratios for semileptonic $B$ decays agree
well with experimental data given by the Particle Data Group
\cite{hik92} or more recently by the CLEO collaboration~\cite{san93}
and ARGUS collaboration~\cite{alb93}. They are given in
Table~\ref{Bbranching}.

\begin{figure}[t]
\vspace{18cm}
\includegraphics{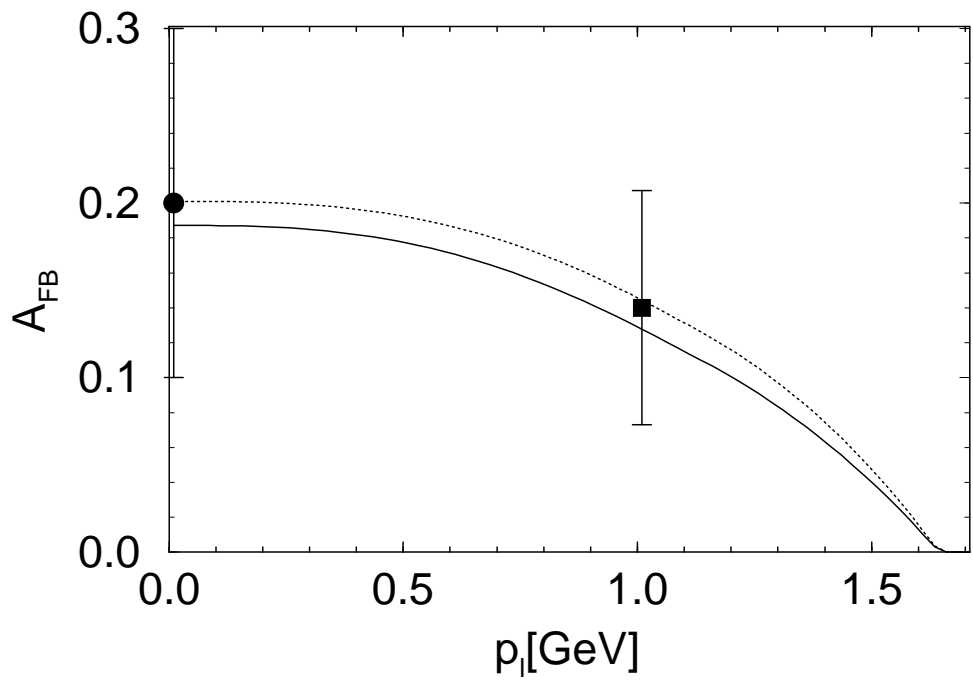}
\vspace{-8cm}
\caption{\label{afb}
Forward backward asymmetry  for $B^0 \rightarrow D^{*+}\ell^-{\ovl
  \nu}$ as a function of lepton momentum 
cut  experiments ARGUS (circle), CLEO  (square) }
\end{figure}

\begin{figure}[t]
\vspace{18cm}
\includegraphics{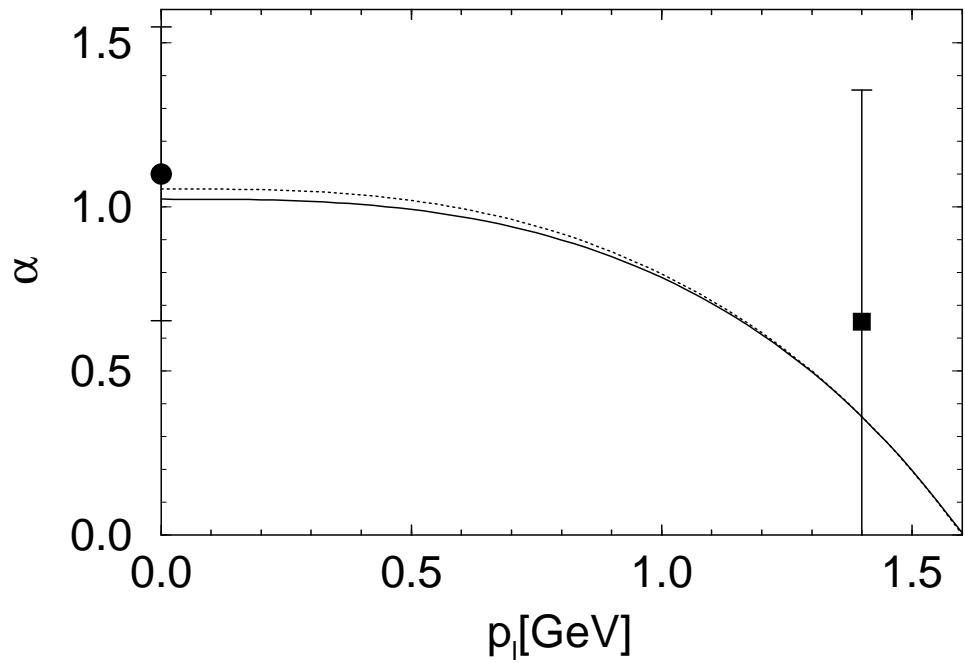}
\vspace{-8cm}
\caption{\label{alpha}
 Asymmetry parameter for 
 $B^0 \rightarrow D^{*+}\ell^-{\ovl \nu}$as function of lepton momentum 
 cut experiments ARGUS (circle), CLEO  (square)}
\end{figure}

In Figs.~\ref{afb} and~\ref{alpha} we give the resulting forward
backward asymmetry $A_{FB}$ and the asymmetry parameter $\alpha$
defined e.g. in~\cite{koe88} as a function of the lepton cut-off
momentum. The forward backward asymmetry sensitive to
parity violation is defined through 
\beq
A_{FB} = \frac{N_F-N_B}{N_F+N_B}
\eeq
with $N$ the number of leptons in forward, resp. backward hemisphere in
the rest system of the $W$-boson. For $A_{FB}$ we have used a
symmetric cut on the lepton momentum $p_\ell$ as utilized by the CLEO
collaboration~\cite{san93} for technical reasons. 

The helicity alignment $\alpha$ describes the $D^{*+}$ polarization
extracted from the $D^{*+}\rightarrow D^0\pi^+$ decay angle
distribution $W(\theta^*)$, viz.
\beq
W(\theta^*)\propto 1+\alpha \cos^2\theta^*
\eeq
For $\alpha$ only a lower cut has been introduced. Both observables
are in good agreement with experimental results.  For comparison we
have included the results of K\"orner-Schuler given in~\cite{koe88} as
a dotted curve.

We would now like to connect our results to the
notion of heavy quark symmetry. In this context form factors
$h_{\pm}(\omega)$, and
$h_V(\omega)$, $h_{A_{1,2,3}}(\omega)$ are introduced. They are
related to the ones introduced above for $ 0^-\rightarrow 0^-$
via \bqn
h(\omega)_\pm&=&\frac{m_B\pm m_D}{2\sqrt{m_Bm_D}} F_1(q^2)
             +\frac{m_B\mp m_D}{2\sqrt{m_Bm_D}} \frac{m_B^2 - m_D^2}{q^2} 
                  \left( F_0(q^2) - F_1(q^2)\right) 
\eqn
and for $0^-\rightarrow 1^-$ via
\bqn
h(\omega)_V&=&\frac{2\sqrt{m_Bm_{D^*}}}{m_B+m_{D^*}} V(q^2)\\
h(\omega)_{A_1}&=&\frac{m_B+m_{D^*}}{(\omega+1)\sqrt{m_Bm_{D^*}}} A_1(q^2)\\
h(\omega)_{A_2}&=
      &\sqrt{\frac{m_B}{m_{D^*}}}\left(A_+(q^2) - A_0(q^2)\right)\\
h(\omega)_{A_3}&=
      &\sqrt{\frac{m_{D^*}}{m_B}}\left(A_-(q^2) + A_0(q^2)\right)
\eqn
where we have introduced  $A_\pm(q^2)$ for convinience.
\beq
A_\pm(q^2) = \left(\frac{m_B}{m_B+m_{D^*}}
              \pm\frac{m_B(m_B+m_{D^*})}{q^2}\right) A_1(q^2)
             \mp \frac{m_B(m_B-m_{D^*})}{q^2} A_2(q^2)
\eeq
In case of
ideal heavy quark limit,
\bqn
h_V(\omega)=h_{A_1}(\omega)=h_{A_3}(\omega)=h_+(\omega)&=&\xi(\omega)\\
h_{A_2}(\omega)=h_-(\omega)&=&0
\eqn
with
\bqn
\omega & = & \frac{m_B^2 + m_{D^{(*)}}^2 - q^2}{2 m_B m_{D^{(*)}}}
\eqn

Note, that we have not employed the limit $m_{D(*)}, m_B \rightarrow
\infty$. The wave functions reproduce the meson mass spectrum and
therefore the overlap at $q^2=q^2_{max}$, viz.  $\omega=1$, cannot be
expected to be complete. Nevertheless, the ideal limit of $\xi(1)=1$
is realized within a few per cent, see Table~\ref{fit}.  We find it
possible to fit the $\omega$ dependence of the dominant form
factor by a simple function, viz.
\beq
h(\omega)=h(1)\cdot\left(1+\beta(1-\omega)\right)
\eeq
with the parameters given in Table~\ref{fit}. The deviations from the
fit are smaller than 1\%, viz. smaller than the model uncertainties
expected.

The form factor $h_-(\omega)$ is small as expected from heavy quark
limit. The form factor $h_{A_2}(\omega)$ has been multiplied by
$r=m_{D^*}/m_B\simeq 0.38$ since this is the relevant quantity
entering in the helicity amplitudes. This way the magnitude of the
form factor $rh_{A_2}(\omega)$ can directly be compared to
$h_{A_3}(\omega)$.

It seems that the heavy quark limit is fulfilled by the model within
10 per cent. Our model is in agreement with the conclusion that the most
sensitivity to mass breaking effects might be expected from the form
factors $h_{A_2}(\omega)$ and $h_V(\omega)$. The form factor
$h_{A_1}(1)$ is close to one, as implied by Luke's
theorem~\cite{luk90}.
 
Concluding this section we now briefly summarize our main results for
heavy to light transitions.

Note that the subject of $\eta$, $\eta^\prime$ mixing is not touched.
Such mixing may be generated in a natural way though instanton effects
\cite{tho76}, or two gluon exchanges. Since none of the above forces
have been considered here, decays into $\eta$ or $\eta^\prime$ are
excluded in the calculation.  At the present stage it would require
additional assumptions and the introduction of mixing parameters, which we
like to exclude here.

Concerning pure leptonic decays of light mesons, the description fails
badly in a nonrelativistic treatment. In particlar, the value of the
pion decay constant $f_\pi$ turns out to be much too large.  Even with
the inclusion of the above mentioned relativistic effects in the
current, the framework does not lead to a satisfactory result. Still,
it is possible to reproduce the pion mass in the framework given here,
see Fig.~\ref{spectra}. Obviously a consistent description of mass
spectra and decay observables of light mesons can not be achieved in a
nonrelativistic framework, compare~\cite{god85} who introduced 'mock'
mesons with mass $\mu=m_q+m_{\ovl q}$ to describe decay data. Some
progress has been achieved using the Bethe Salpeter equation for light
$q{\ovl q}$ systems, and spectra as well as decays of light mesons can
be described~\cite{mue93}.

With this in mind, one has to be careful in interpreting decays
involving (very) light mesons in the final state.

The form factors, viz. $B(D^{(*)}) \rightarrow K,K^*,\pi,\rho, \omega$
and $D_s\rightarrow K, K^*, \phi$ differ from the pole dominance
hypothesis at $q^2=q^2_{max}$, viz. small three momentum
transfer~\cite{res93}. 
However, they lead to similar values as in the relativistic harmonic
oscillator model at $q^2=0$~\cite{bau87}.

Where experimental data is given, we compare our results of the total
branching ratios of $D$-meson decays and find a rather good agreement.
Results are shown in Table~\ref{Dbranching}.

\section{Nonleptonic decays}

Nonleptonic weak decays provide additional phenomena that are connected to
 strong interaction. Examples are hard gluon exchanges,
quark rearrangement, annihilation and long range effects. Thus,
extraction of fundamental physical constants such as the
Cabbibo-Kobayashi-Maskawa matrix elements is more difficult than in
the semileptonic case.

Following~\cite{koe88} we introduce the effective Lagrangian for
$\Delta B=1$ transitions including QCD-effects but neglecting
pinguin contributions~\cite{shi79},
\bqn
\label{efflagrangian}
\CL_{eff} & = & -\frac{G_F}{\sqrt{2}} \sum_{\alpha^\prime \alpha\beta}
 V_{\alpha^\prime b} V^*_{\alpha\beta}  
\left( C_1(m)h_{\mu,\alpha\beta}h^\mu_{\alpha^\prime b}
+ C_2(m)h_{\mu,\alpha^\prime \beta }h^\mu_{\alpha b} \right)
\eqn

with $\alpha,\alpha^\prime \in\{u,c\}$, $\beta\in\{d,s\}$, and the
Wilson coefficients $C_1(m)$, $C_2(m)$ depending on the scale $m$.
For the scale $m_c\simeq 1.5$GeV one expects $C_1(m_c)=1.27$ and
$C_2(m_c)=-0.53$, and for $m_b\simeq 5$GeV the resulting values are
$C_1(m_b)= 1.12$ and $C_2(m_b)= -0.26$~\cite{alt74}. Note that if no
additional gluon exchanges are assumed in the operator, the Lagrangian
(\ref{efflagrangian}) reduces to the nonleptonic weak Langrangian in
current current approximation since then $C_1(m)=1$ and $C_2(m)=0$.

The Lagranian (\ref{efflagrangian}) is evaluated between the meson
amplitudes given in (\ref{wavefct}). If final state interaction
($fsi$) is neglected, the transition amplitude factorizes.  Due to
Fierz rearrangement two generic types of contributions are possible.
The generic form of class I is given by (\ref{hypfactor}) those of
class II by (\ref{ahypfactor}), class III are mixed forms of both e.g.
$B^- \rightarrow D^0\pi^-$. 
\bqn
\bramket{\pi^+\;D^-}{h_{\mu,ud} h^\mu_{cb}} {B^0} 
&\rightarrow
&a_1\bramket{\pi^+}{h_{\mu,ud}}{0} \bramket{D^-}{h^\mu_{cb}}{B^0}
\label{hypfactor}\\
\bramket{\pi^0\;D^0}{h_{\mu,ud}h^\mu_{cb}}{B^0} 
&\rightarrow
&a_2\bramket{D^0}{h_{\mu,cd}}{0}\bramket{\pi^0}{h^\mu_{ub}}{B^0}
\label{ahypfactor}
\eqn
where we have introduced $a_1$ and $a_2$, to connect our results with
the model of BSW~\cite{bau87}. 
In the above case $a_1=1$ and from Fierz rearrangement $a_2=1/3$.  If
gluonic contributions are taken into account, $a_1=C_1(m)+ \xi C_2(m)$
and $a_2 = C_2(m) + \xi C_1(m)$, with $\xi=1/3$.

In fact BSW have introduced $a_1$ and $a_2$ on the level of the
effective Lagrangian and assume $h_{\alpha \beta}$ to depend only on
asymptotic hadronic (interpolating) fields. This introduces problems
in interpreting $a_1$ and $a_2$ in terms of the Wilson coefficients
and therefore $a_1$ and $a_2$ are usually treated as free parameters,
see discussion in~\cite{bau87} and references therein. Since we have
introduced wave functions for the mesons, which connect quark degrees
of freedom with asymptotic degrees of freedom these parameters have a
different interpretation as in BSW. However, the general problems
discussed in \cite{bau87} remain the same, and $a_1$ and $a_2$ can be
treated as free parameters in the wave function approach also.  Since
the same approximations are done when calculating matrix elements, our
values of $a_1$ and $a_2$ are compatible with those given by BSW.

Evaluation of the r.h.s. of (\ref{hypfactor}) and (\ref{ahypfactor})
using quark model wave functions leads to some uncertainty connected
to light mesons.  As already mentioned in the previous section light
mesons are not well described. Part of the uncertainty related e.g.
to the transition matrix element $\bramket{X}{h_\mu}{0}$ can be
removed by using empirical decay constants instead of calculated ones.
From leptonic weak decays these are available only for $\pi$ and $K$
decays. For the $D$-meson an upper limit exists, $f_{D^+}<310$MeV.
For $f_{D_s}$ we have chosen $f_{D_s}=300$MeV, which is a good fit of
our model to the branching ratios measured (see Table~\ref{BDdecays}),
and is close to values found by Rosner~\cite{ros90} and an ARGUS
analysis using different type of model analyses~\cite{alb92}. For
$f_{\rho}$ we have used $f_\rho=205$MeV, which has been suggested
by~\cite{neu92}. Other decay constants are taken from~\cite{bau87} and
listed in Table~\ref{decayconstants}.
 
The problem of factorization usually discussed in this context is not
further elaborated here, see~\cite{bau87, bjo89,bur86}. Through the
use of wave functions many important questions related to the
factorization hypothesis have to be rephrased (e.g.  color octet
excitations vs. model space etc.).  Model independent tests of
factorization have been suggested by Kamal, Xu and Czarnecki, who
found that even in cases of heavy to light-light (except
$D^0\rightarrow K^- a_1^+$), factorization seems reasonable, provided
the final state interaction is properly taken into
account~\cite{kam93}.

We consider the following combinations of decays $0^- \rightarrow 0^-
(0^-)$, $0^- \rightarrow 0^- (1^-)$, $0^- \rightarrow 0^- (1^+)$, $0^-
\rightarrow 1^- (0^-)$, $0^- \rightarrow 1^- (1^-)$, and $0^-
\rightarrow 1^- (1^+)$. The quantum numbers of the meson due to vacuum
to meson transition is given in parenthesis. Experimental data are
taken from the compilation of the Particle Data Group~\cite{hik92}.
However, were new data exsist, we have updated the average values
using the latest results of CLEO~\cite{san93} and ARGUS~\cite{alb92,
  alb93} collaborations.

In each of the following tables~\ref{BDdecays}-\ref{BDrestdecay} we
have given the type of decay, our calculated decay rate excluding CKM
matrix elements and decay constants, our decay rate in terms of $a_1$
and/or $a_2$, which may be compared with the respective rates of
BSW~\cite{bau87,neu92}, and in the last colmun the experimental value
where available. The decay parameters used are quoted in
Table~\ref{decayconstants}. The Cabbibo Kobayashi Maskawa matrix
elements used are given in Table~\ref{CKMparameters}.

From $ {\ovl B^0} \rightarrow D^+ \pi^-$, $ {\ovl B^0} \rightarrow D^+
\rho^-$, $ {\ovl B^0} \rightarrow D^{*+} \pi^-$, and $ {\ovl B^0}
\rightarrow D^{*+} \rho^-$ decays we find a best fit to the data for
$a_1 = (0.96\pm 0.05)(0.036/V_{cb})(1.32ps/\tau_B)^{1/2}$. Inclusion
of all decays in $\chi^2$ fitting leads to essentially no change
on $a_1$.

For $a_2$ we find $|a_2|=0.31\pm 0.03$ as a best fit to the data.
Including recent CLEO data, we find the relative sign between $a_1$
and $a_2$ positive in the following sense. The total $\chi^2$ from the
decays $B^-\rightarrow D^0 \pi^-$, $B^-\rightarrow D^{*0} \pi^-$,
$B^-\rightarrow D^0 \rho^-$, and $B^-\rightarrow D^{*0} \rho^-$ is
$\chi^2_+\simeq 2.6$ for the plus sign and for the minus sign
$\chi^2_- \simeq 25$. Details are given in Table~\ref{BDrestdecay}.

Some experimental uncertainties may be eliminated using ratios of
branching ratios, and with the assumtion $f_{D_s}=f_{D^*_s}$ less
ambiguous comparisons can be made. In table~\ref{superratios} we have
compared our results to recent ARGUS data~\cite{alb92}.

\section{Summary and Conclusion}

We have calculated ${\ovl q}q$ spectra within a nonrelativistic
framework with forces expanded up to order $(p/m)^2$. Not all terms in
the Hamiltonian are equally important. Those of minor importance are
left out for convenience. We find parameter values in a reasonable
range and close to those of the Cornell potential (where comparison is
possible). Spectra for all mesons are reproduced reasonably well. We
have, however, excluded the question of $\eta$, $\eta^\prime$ mixing,
which is not expected to influence the conclusion reached in this
section.

We have included the full Dirac quark spinor to define the appropriate
current operators for the meson Fock space.  Compared to the
nonrelativistic approach we found important effects on decay constants
and form factors.

Since we use physical masses, the Isgur Wise function cannot be
calculated. However, we find that at $\omega=1$ the normalization of
the form factors is within $10\%$ of the heavy quark limit
expectations.

Assuming the lifetime of $B$-meson $\tau_B=1.32ps$, which is longer
than given in the last edition of partical data tables~\cite{hik92},
but still shorter than most recent values, we find $V_{cb}=0.036\pm
0.003$. 

Concerning nonleptonic decays, the best values for $a_1$ and $a_2$
suggest that hard gluon exchanges may be neglected, viz.
$C_2(m_b)\simeq 0$. A similar conclusion is possible for charmonium
and bottomonium within the framework of the quark model presented
here, see discussion in~\cite{bey92}. Gluonic effects have been found
to accommodate experimental data only in a pure nonrelativistic
approach of lowest order.  However, if relativity is treated more
appropriately, inclusion of gluonic effects destroy the rather good
agreement with quarkonium data~\cite{bey92}.

Other suggestions are also possible. For example, if the color factor
$\xi$ introduced after (\ref{ahypfactor}) were changed to $\xi \simeq
1/2$, then $a_1=0.99$ and $a_2=0.30$ at the $B$ meson mass scale. These
numbers are also in reasonable agreement with the value extracted
here. This might imply that `color octett' contributions are not
negligible. 

However, at the present stage of analysis we would like to
emphasize that the above possibilities are mere speculations, which
however call for further investigation. I seems that the situation has
never been as puzzling as at the moment. Since relativistic effects
seem to play a nonnegligible role, some progress can be expected from
recent developements using Bethe-Salpeter equation taking covariance
seriously.

We have not considered nonleptonic $D$ decays in the paper. No
reasonable $\chi^2$ could be found for a fit of $a_1$ and $a_2$. It is
well known that these decays are more biased by $fsi$ and relativistic
effects, which we did not take into account. Also the model (being
generically nonrelativistic) should be less reliable in these cases.

\newpage

\begin{table}
\caption{\label{parameter}
Parameter values of the Hamiltonian. 
 $^*$For  mesons other than $B$-, and $D^{(*)}_{(s)}$-mesons 
we have used $r_0=0.30$fm}
\[
\begin{array}{|l|c|}
\hline
m_n {\rm ~[GeV]}&0.411 \\[0.5ex]
m_s {\rm ~[GeV]} &0.594 \\[0.5ex]
m_c{\rm  ~[GeV]}& 1.806 \\[0.5ex]
m_b {\rm ~[GeV]}&5.183 \\[0.5ex]
a {\rm   ~[GeV]} &-0.668 \\[0.5ex]
b {\rm   ~[GeV/fm]} &0.792 \\[0.5ex]
\alpha_s &0.41 \\[0.5ex]
r_0 {\rm ~[fm]}^*&0.24\\[0.5ex]\hline
\end{array}
\]
\end{table}

\begin{table}
\caption{\label{polmasses}
Pole masses (in GeV) for $B \rightarrow D$ and $B 
\rightarrow D^*$ form factors }
\[
\begin{array}{|cccc|}
\hline
F_0  & F_1, V &  A_1, A_2  &  A_0 \\ \hline\hline
6.80 & 6.34 & 6.37 & 6.3 \\ \hline
\end{array}
\]
\end{table}

\begin{table}
\caption{\label{Bbranching}
Branching ratios for B decays. We use $\tau_B =1.32$ps, a)
recent CLEO, b) recent ARGUS data}
\[
\begin{array}{|r@{}c@{}l|cc|}
\hline
\multicolumn{3}{|c|}{\makebox{decay}}&Br_{QM} [\%]&Br_{exp} [\%]\\
\hline\hline
{\ovl B}^0 &\rightarrow  &D^{*+}\ell^-{\ovl \nu}_\ell &4.5 &4.9\pm 0.8\\
&&&&4.50\pm 0.44 \pm 0.44^a\\
&&&&5.2\pm 0.5\pm 0.6^b\\
{\ovl B}^0 &\rightarrow  &D^{+}\ell^-{\ovl \nu}_\ell  &1.8 &1.6 \pm 0.7\\ 
\hline
\end{array}
\]
\end{table}

\begin{table}
\caption{\label{fit}
Parameters for a fit to the dominant formfactors}
\[
\begin{array}{|c|c|ccc|}
\hline
&h_+&h_V&h_{A_1}&h_{A_3}\\
\hline\hline
h(1)& 0.993 & 0.896   &  0.977      &  0.945\\
\beta& 0.47  & 0.58 & 0.58 & 0.58 \\
\hline
\end{array}
\]
\end{table}

\begin{table}
\caption{\label{Dbranching}
Branching ratios for D decays. We use 
$\tau_{D^0} =0.42$ps, $\tau_{D^+} =1.066$ps, $\tau_{D_s} =0.45$ps}
\[
\begin{array}{|r@{}c@{}l|cc|}
\hline
\multicolumn{3}{|c|}{\makebox{decay}}&Br_{QM} [\%]&Br_{exp} [\%]\\
\hline\hline
D^+ &\rightarrow  &{\ovl K}^0  e^+{\ovl \nu}_e     &7.64 &5.5 \pm 1.2\\
D^+ &\rightarrow  &{\ovl K}^0  \mu^+{\ovl \nu}_\mu &7.64 &7.0 \pm 3.0\\
D^+ &\rightarrow  &{\ovl K^*}^0e^+{\ovl \nu}_e     &6.43 &4.1 \pm 0.6\\
D^0 &\rightarrow  &K^-   e^+  {\ovl \nu}_e         &3.01 &3.31\pm 0.29\\
D^0 &\rightarrow  &K^-   \mu^+{\ovl \nu}_\mu       &3.01 &2.9 \pm 0.5\\
D^0 &\rightarrow  &K^{*-}e^+ {\ovl \nu}_e          &2.53 &1.7 \pm 0.6\\
D^s &\rightarrow  &\phi  \ell^+ {\ovl \nu}_\ell    &2.41 &1.4 \pm 0.5\\
\hline
\end{array}
\]
\end{table}

\begin{table}
\caption{\label{decayconstants}
Leptonic decay constants used in the calculation }
\vspace{0.5cm}
\centering
\begin{tabular}{|ccc|ccc|}
\hline
weak       & meson      & $f$  & weak  & meson       & $f$  \\
current    & type           & [MeV] & current & type   & [MeV] \\
 \hline\hline
$\bar u d$ & $\pi^-$    & 132 & $\bar u d$ & $\rho^-$      & 205 \\
$\bar d d$ & $\pi^0$    & 93  & $\bar d d$ & $\rho^0$      & 145 \\
$\bar u s$ & $K^-$      & 162 & $\bar u d$ & $K^{*-}$      & 220 \\
$\bar d s$ & $\bar K^0$ & 162 & $\bar u d$ & $\bar K^{*0}$ & 220 \\
$\bar d c$ & $D^+$      & 220 & $\bar d c$ & $D^{*+}     $ & 220 \\
$\bar u c$ & $D^0$      & 220 & $\bar u c$ & $D^{*0}$      & 220 \\
$\bar s c$ & $D_s^+$    & 300 & $\bar s c$ & $D_s^{*+}$    & 300 \\
           &            &     & $\bar d d$ & $\omega$      & 145 \\
           &            &     & $\bar c c$ & $J/\Psi$      & 382 \\
           &            &     & $\bar u d$ & $a_1^-$       & 220 \\
\hline
\end{tabular}
\end{table}

\begin{table}
\caption{\label{CKMparameters}
Parameters used in the calculation of non-leptonic
 decays}
\vspace{0.5cm}
\centering
\begin{tabular}{|cccccc|}
\hline
$V_{u d}$&$V_{u s}$ &$V_{u b}$ &$V_{c d}$ &$V_{c s}$
&$V_{c b}$\\[0.5ex]
\hline\hline
0.9753   &0.221    &-         &0.221     &0.9743    &0.036 \\
\hline
\end{tabular}
\end{table}

\begin{table}
\renewcommand{\arraystretch}{1.3}
\caption{\label{superratios}
Comparison of ratios of branching ratios, upper part
  ARGUS data~\protect\cite{alb92}}
\vspace{0.5cm}
\[
\begin{array}{|cccc|}
\hline
 \frac{Br(B^-\rightarrow D^{*-}_s D^0)}   
       {Br(B^-\rightarrow D^{-}_s  D^0)}
&\frac{Br(B^-\rightarrow D^{*-}_s D^{*0})}
       {Br(B^-\rightarrow D^{-}_s  D^{*0})}
&\frac{Br(B^0\rightarrow D^{*-}_s D^+)}
       {Br(B^0\rightarrow D^{-}_s  D^+)}
&\frac{Br(B^0\rightarrow D^{*-}_s D^{*+})}
       {Br(B^0\rightarrow D^{-}_s  D^{*+})}\\
\hline\hline
0.67\pm 0.65 & 2.4 \pm 2.1&  1.6 \pm 1.5  & 1.9\pm 1.6 \\
0.63&3.3&0.63&3.3\\
\hline
\end{array}
\]
\end{table}

\begin{table}
\renewcommand{\arraystretch}{1.3}
\caption{\label{BDdecays}
Class I $B$ decay and branching ratios
due to $ \bar c b$ current. Model calculation with  
$a_1=0.96 (0.036/V_{cb})(1.32ps/\tau_B)^{1/2}$, for 
normalization of $B\rightarrow D^{(*)}D^{(*)}_s$ decays 
we have used
$Br(D_s^+\rightarrow \phi\pi^+)=2.7\%$ as suggested by ARGUS. }
\vspace{0.5cm}
\centering
\begin{tabular}{|l|c@{}c@{}cc|c|}
\hline
decay mode &\multicolumn{4}{c|}{\makebox{quark model}} 
&experiment \\
& $\Gamma [10^8 s^{-1}]$&
&Br[\%] &Br[\%]  &Br[\%]\\
\hline\hline
$ {\ovl B}^0 \rightarrow D^+ \pi^-$ & $1.233 V_{cb}^2 V_{u d}^2
    f_\pi^2 a_1^2$ & $\rightarrow$ &
0.350 $a_1^2$ & 0.322& $0.29 \pm 0.05$   \\
${\ovl B}^0 \rightarrow D^+ \rho^-$ & $1.162 V_{cb}^2 V_{u d}^2
    f_\rho^2 a_1^2$ & $\rightarrow$ &
0.795 $a_1^2$&  0.732 & $0.73 \pm 0.21$  \\
${\ovl B}^0 \rightarrow D^+ a_1^-$ & $1.038  V_{cb}^2 V_{u d}^2
    f_{a_1}^2 a_1^2$ & $\rightarrow$ &
0.818 $a_1^2$ &  0.753 & $0.60 \pm 0.33$  \\
$ {\ovl B}^0 \rightarrow D^{*+} \pi^-$ & $0.953 V_{cb}^2 V_{u d}^2
    f_\pi^2 a_1^2$ & $\rightarrow$ &
0.270 $a_1^2$ &  0.249  & $0.30 \pm 0.05$ \\
${\ovl B}^0 \rightarrow D^{*+} \rho^-$ & $1.093 V_{cb}^2 V_{u d}^2
    f_\rho^2 a_1^2$ & $\rightarrow$ &
0.747 $a_1^2$ & 0.689  & $0.74 \pm 0.17$  \\
${\ovl B}^0 \rightarrow D^{*+} a_1^-$ & $1.304  V_{cb}^2 V_{u d}^2
    f_{a_1}^2 a_1^2$ & $\rightarrow$ &
1.027 $a_1^2$ & 0.947  & $1.80 \pm 0.85$  \\
${\ovl B}^0 \rightarrow D^{+} D_s^-$ & $1.072  V_{cb}^2 V_{c s}^2
    f_{D_s}^2 a_1^2$ & $\rightarrow$ &
1.567 $a_1^2$  & 1.444  & $0.84 \pm 0.51$ \\
${\ovl B}^0 \rightarrow D^{+} D_s^{*-}$ & $0.674 V_{cb}^2 V_{c s}^2
    f_{D_s^*}^2 a_1^2$ & $\rightarrow$ &
0.985 $a_1^2$  & 0.908 & $2.7\pm 1.9$  \\
${\ovl B}^0 \rightarrow D^{*+} D_s^-$ & $0.524 V_{cb}^2 V_{c s}^2
    f_{D_s}^2 a_1^2$ & $\rightarrow$ &
0.765 $a_1^2$ &0.705  & $1.42 \pm 0.75$  \\
${\ovl B}^0 \rightarrow D^{*+} D_s^{*-}$ & $1.721V_{cb}^2 V_{c s}^2
    f_{D_s^*}^2 a_1^2$ & $\rightarrow$ &
2.515 $a_1^2$ & 2.318 & $2.6\pm 1.5$   \\
${\ovl B}^0 \rightarrow D^{+} K^-$ & $1.227 V_{cb}^2 V_{u s}^2
    f_{K}^2 a_1^2$ & $\rightarrow$ &
0.027 $a_1^2$ & 0.025 & ---  \\
${\ovl B}^0 \rightarrow D^{+} K^{*-}$ & $1.137 V_{cb}^2 V_{u s}^2
    f_{K^*}^2 a_1^2$ & $\rightarrow$ &
0.046 $a_1^2$  & 0.042& ---  \\
${\ovl B}^0 \rightarrow D^{*+} K^-$ & $0.928 V_{cb}^2 V_{u s}^2
    f_{K}^2 a_1^2$ & $\rightarrow$ &
0.020 $a_1^2$  &  0.019 & --- \\
${\ovl B}^0 \rightarrow D^{*+} K^{*-}$ & $1.138 V_{cb}^2 V_{u s}^2
    f_{K^*}^2 a_1^2$ & $\rightarrow$ &
0.046 $a_1^2$ &0.042  & ---  \\
$ {\ovl B}^0 \rightarrow D^+ D^-$ & $1.093  V_{cb}^2 V_{c d}^2
    f_D^2 a_1^2$ & $\rightarrow$ &
0.044 $a_1^2$  & 0.041 & --- \\
$ {\ovl B}^0 \rightarrow D^+ D^{*-}$ & $0.726 V_{cb}^2 V_{c d}^2
    f_{D^*}^2a_1^2$ & $\rightarrow$ &
0.029 $a_1^2$ & 0.027 & ---  \\
$ {\ovl B}^0 \rightarrow D^{*+} D^-$ & $0.565 V_{cb}^2 V_{c d}^2
    f_D^2 a_1^2$ & $\rightarrow$ &
0.023 $a_1^2$  & 0.021  & --- \\
$ {\ovl B}^0 \rightarrow D^{*+} D^{*-}$ & $1.682 V_{cb}^2 
V_{c d}^2
    f_{D^*}^2 a_1^2$ & $\rightarrow$ &
0.068 $a_1^2$  & 0.063 & --- \\[0.5cm]
$ B^- \rightarrow D^0 a_1^-$ & $1.038  V_{cb}^2 V_{u d}^2
    f_{a_1}^2 a_1^2$ & $\rightarrow$ &
0.818 $a_1^2$ &  0.753  & $0.45 \pm 0.36$ \\
$ B^- \rightarrow D^{*0} a_1^-$ & $1.304  V_{cb}^2 V_{u d}^2
    f_{a_1}^2 a_1^2$ & $\rightarrow$ &
1.027 $a_1^2$  & 0.947  & ---\\
$ B^- \rightarrow D^{0} D_s^-$ & $1.072  V_{cb}^2 V_{c s}^2
    f_{D_s}^2 a_1^2$ & $\rightarrow$ &
1.567 $a_1^2$  &  1.444 & $2.0 \pm 0.8$ \\
${\ovl B}^- \rightarrow D^{0} D_s^{*-}$ & $0.674 V_{cb}^2 V_{c s}^2
    f_{D_s^*}^2 a_1^2$ & $\rightarrow$ &
0.985 $a_1^2$  &   0.908& $1.6\pm 1.2$ \\
${\ovl B}^- \rightarrow D^{*0} D_s^-$ & $0.524 V_{cb}^2 V_{c s}^2
    f_{D_s}^2 a_1^2$ & $\rightarrow$ &
0.765 $a_1^2$ & 0.705  & $1.3 \pm 0.9$ \\
${\ovl B}^- \rightarrow D^{*0} D_s^{*-}$ & $1.72 
V_{cb}^2 V_{c s}^2   f_{D_s^*}^2 a_1^2$ & $\rightarrow$ &
2.514 $a_1^2$  &  2.317 & $3.1 \pm 1.7$  \\
$ {\ovl B}^- \rightarrow D^0 D^-$ & $1.093 V_{cb}^2 V_{c d}^2
    f_D^2 a_1^2$ & $\rightarrow$ &
0.044 $a_1^2$  &  0.041 & --- \\
$ {\ovl B}^- \rightarrow D^0 D^{*-}$ & $0.726 V_{cb}^2 V_{c d}^2
    f_{D^*}^2a_1^2$ & $\rightarrow$ &
0.029 $a_1^2$ &  0.027 & --- \\
$ {\ovl B}^- \rightarrow D^{*0} D^-$ & $0.565 V_{cb}^2 V_{c d}^2
    f_D^2 a_1^2$ & $\rightarrow$ &
0.013 $a_1^2$  &  0.021 & --- \\
\hline
\end{tabular}
\end{table}

\begin{table}
\renewcommand{\arraystretch}{1.3}
\caption{\label{ubdecay}
Class I $B$ decay and branching ratios due to 
$ \bar u b$ current, ${\tilde a}_1=a_1 
(V_{bu}/V_{bc})$ }
\vspace{0.5cm}
\centering
\begin{tabular}{|l|c@{}c@{}cc|c|}
\hline
decay mode &\multicolumn{4}{c|}{\makebox{quark model}} 
&experiment \\
& $\Gamma [10^8 s^{-1}]$&
&Br[\%] &Br[\%]  &Br[\%]\\
\hline\hline
$ {\ovl B}^0 \rightarrow \pi^+ \pi^-$ & $0.170 V_{u b}^2 V_{u d}^2
    f_\pi^2 a_1^2$ & $\rightarrow$ & 0.048 ${\tilde a}_1^2$
 &  0.044 $ (V_{bu}/V_{bc})^2$& $< 0.009$ \\
${\ovl B}^0 \rightarrow \pi^+ \rho^-$ & $0.176 V_{u b}^2 V_{u d}^2
    f_\rho^2 a_1^2$ & $\rightarrow$ & 0.120 ${\tilde a}_1^2$
 &   0.111  $ (V_{bu}/V_{bc})^2$& ---\\
${\ovl B}^0 \rightarrow \pi^+ a_1^-$ & $0.185 V_{u b}^2 V_{u d}^2
 f_{a_1}^2 a_1^2$ & $\rightarrow$ & 0.146 ${\tilde a}_1^2$
 & 0.136  $ (V_{ub}/V_{bc})^2$& $<0.057$ \\
\hline
\end{tabular}
\end{table}

\begin{table}
\renewcommand{\arraystretch}{1.3}
\centering 
\caption{\label{Bdecay}
Class II $B$ decay and branching ratios, $|a_2|=0.31$}
\vspace{0.5cm}
\begin{tabular}{|l|c@{}c@{}cc|c|}
\hline
decay mode  &\multicolumn{4}{c|}{\makebox{quark model}}&experiment\\
& $\Gamma [10^8 s^{-1}]$&
&Br[\%] &Br[\%]  &Br[\%]\\
\hline\hline
$ {\ovl B}^0 \rightarrow \pi^0 D^0$ & $0.111 V_{c b}^2 V_{u d}^2
    f_\pi^2 a_2^2$ & $\rightarrow$ & 0.087 $a_2^2 $ &  0.008 & --- \\
${\ovl B}^0 \rightarrow \pi^0 D^{*0}$ & $0.104 V_{u b}^2 V_{u d}^2
    f_{D^*}^2 a_2^2$ & $\rightarrow$ & 0.082 $a_2^2 $ &  0.008 & --- \\
${\ovl B}^0 \rightarrow K^0 J/\Psi$ & $0.320 V_{c b}^2
V_{c s}^2 f_{J/\Psi}^2 a_2^2$ & $\rightarrow$ & 0.759 $a_2^2 $ & 0.073
& $0.077 \pm 0.026$ \\
${\ovl B}^0 \rightarrow K^{*0} J/\Psi$ & $0.732 V_{c b}^2
V_{c s}^2 f_{J/\Psi}^2 a_2^2$ & $\rightarrow$ & 1.735 $a_2^2 $ &
0.167 & $0.14 \pm 0.03$\\
${\ovl B}^0 \rightarrow K^{*0} D^{*0}$ & $0.524 V_{c b}^2
V_{u s}^2 f_{D^*}^2 a_2^2$ & $\rightarrow$ & 0.021 $a_2^2 $ &  0.002
& --- \\
${\ovl B}^0 \rightarrow \pi^0 J/\Psi$ & $0.231  V_{c b}^2
V_{c d}^2 f_{J/\Psi}^2 a_2^2$ & $\rightarrow$ & 0.028 $a_2^2 $ &
0.003 & --- \\[0.3cm]
$ B^- \rightarrow K^- J/\Psi$ & $0.320 V_{c b}^2
V_{c s}^2 f_{J/\Psi}^2 a_2^2$ & $\rightarrow$ & 0.759 $a_2^2 $ & 0.073
& $0.090 \pm 0.014$ \\
$ B^- \rightarrow K^{*-} J/\Psi$ & $0.732  V_{c b}^2
V_{c s}^2 f_{J/\Psi}^2 a_2^2$ & $\rightarrow$ & 1.735 $a_2^2 $&  0.167
& $0.16 \pm 0.05$  \\
$ B^- \rightarrow \pi^- J/\Psi$ & $0.463 V_{c b}^2
V_{cd}^2 f_{J/\Psi}^2 a_2^2$ & $\rightarrow$ & 0.056 $a_2^2 $  &
0.005 & --- \\
\hline
\end{tabular}
\end{table}

\begin{table}
\renewcommand{\arraystretch}{1.3}
\centering 
\caption{\label{BDrestdecay}
Class III (mixed) $B$ branching ratios}
\vspace{0.5cm}
\begin{tabular}{|l|ccc|c|}
\hline
decay mode   &\multicolumn{3}{c|}{\makebox{quark model Br[\%]}}
&experiment\\
& &$a_2>0$ &$a_2<0$& Br[\%]\\
\hline\hline
$B^- \rightarrow D^0 \pi^-$ & 
0.344 $(a_1+0.52 a_2)^2$&  0.433 &0.220
&$0.38\pm 0.05$ \\
$B^- \rightarrow D^{*0} \pi^-$ & 
0.266 $(a_1+0.80 a_2)^2$ &   0.388 &0.135
&$0.42\pm 0.10$\\
$B^- \rightarrow D^0 \rho^-$ & 
0.795 $(a_1+0.55 a_2)^2$&  1.016& 0.496
&$1.08\pm 0.27$ \\
$B^- \rightarrow D^{*0} \rho^-$ & 
1.152 $(a_1+0.59 a_2)^2$&   1.507 &0.694
&$1.11\pm 0.35$\\
$B^- \rightarrow D^0 K^-$ & 
0.027 $(a_1+0.55 a_2)^2$ & 0.034 &0.017 & ---  \\
$B^- \rightarrow D^0 K^{*-}$ & 
0.047 $(a_1+0.46 a_2)^2$ &  0.057 & 0.031& --- \\
\hline
\end{tabular}
\end{table}

\newpage

\begin{figure}[t]
\vspace{18cm}
\includegraphics{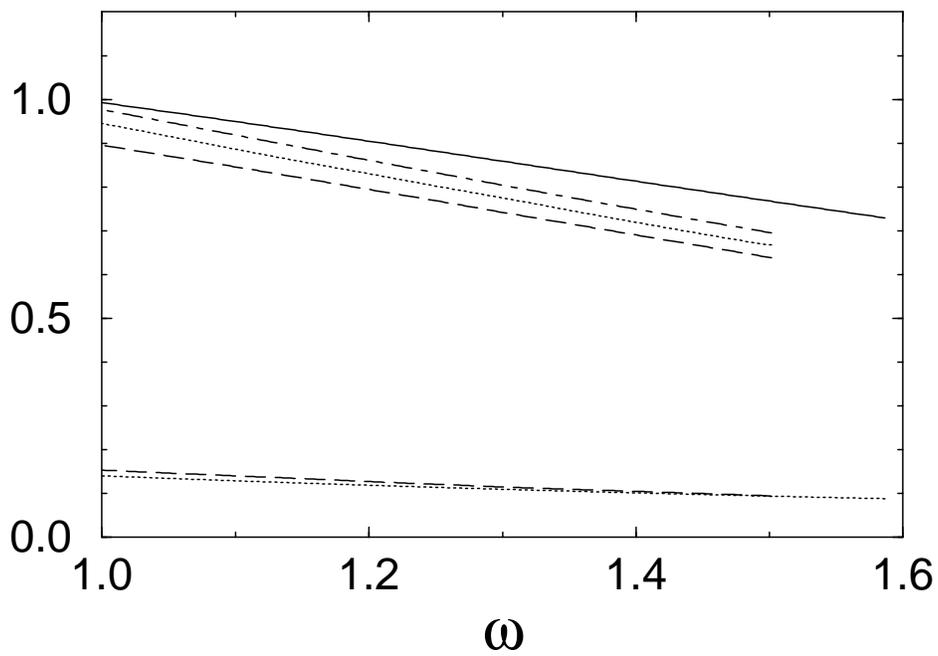}
\vspace{-9cm}
\caption{\label{isgurwise}
  ``Heavy quark'' form factors for  B $\rightarrow$ D und
  B $\rightarrow$ $D^*$ transitions; from top to bottom: $h_+$,
  $h_{A_1}$, $h_{A_3}$, $h_V$, $rh_{A_2}$, $h_-$, $r=m_{D^*}/m_B$.}
\end{figure}

\end{document}